\documentclass[aps,prd,12pt,showpacs,notitlepage,nofootinbib,tightenlines]{revtex4-1}
\usepackage{amsmath}
\usepackage{bm}
\usepackage{times}
\usepackage{braket}
\usepackage{color}
\usepackage{epsfig}
\usepackage{slashed}
\usepackage{hyperref}

\newcommand{\beq}{\begin{eqnarray}}
\newcommand{\eeq}{\end{eqnarray}}
\newcommand{\non}{\nonumber\\ }

\newcommand{\cala}{{\cal A}}

\def \npb{ {\bf Nucl.Phys. B} }
\def \plb{ {\bf Phys.Lett. B} }
\def \pr{  {\bf Phys. Rep.} }
\def \prd{ {\bf Phys.Rev. D} }
\def \prl{ {\bf Phys.Rev.Lett.}  }

\def \jhep{ {\bf J. High Energy Phys.}  }
\definecolor{Red}{rgb}{1.,0.,0.}

\definecolor{Blue}{rgb}{0.,0.,1.}

\definecolor{nicered}{rgb}{0.7,0.1,0.1}
\definecolor{nicegreen}{rgb}{0.1,0.5,0.1}

\hypersetup{colorlinks,citecolor=nicegreen,linkcolor=nicered}

\begin{document}
\title{Left-right twin Higgs model confronted with the latest LHC Higgs data}
\author{Yao-Bei Liu$^{1,2}$, Shan Cheng$^{1}$, Zhen-Jun Xiao$^{1}$ \footnote{Electronic address: xiaozhenjun@njnu.edu.cn}}
\affiliation{1. Department of Physics and Institute of Theoretical Physics,
Nanjing Normal University, Nanjing, Jiangsu 210023, P.R.China \\
  2. Henan Institute of Science and Technology, Xinxiang 453003, P.R.China}
\date{\today}
\begin{abstract}
Motivated by the latest LHC Higgs data, we calculate the new physics contributions
to the Higgs decay channels of $h\to \gamma\gamma, Z\gamma, \tau\tau, WW^*$ and $ ZZ^*$ in the
left-right twin Higgs (LRTH) model, induced by the loops involving the heavy T-quark, the $W_H$ and
$\phi^\pm$ bosons appeared in the LRTH model. We find that
(a) for a SM-like Higgs boson around 125.5 GeV, the signal rates normalized to the corresponding
standard model (SM) predictions are always suppressed when new physics contributions are taken
into account and approach the SM predictions for a large scalar parameter $f$;
and (b) the LRTH prediction for $R_{\gamma\gamma}$  agree well with the CMS measurement
$R_{\gamma\gamma}=0.77\pm 0.27$ at $1\sigma$ level, but differ with the ATLAS result.
The forthcoming precision measurement of the diphoton signal at the LHC
can be a sensitive probe for the LRTH model.
\end{abstract}


\maketitle

\newpage
\section{Introduction}

Very Recently, the discovery of a neutral Higgs boson  at CERN's
Large Hadron Collider (LHC) experiment has been confirmed by the
ATLAS and CMS collaborations
\cite{atlas-1,cms-2,atlas-1b,Roeck-lp2013,atlas-2013,atlas-2013b}.
This discovery is based on the Higgs boson search with a variety of
Higgs boson decay modes. Among the major decay modes of a standard
model (SM) Higgs boson studied intensively at ATLAS and CMS
experiments, the diphoton channel is one of the most important
channels for Higgs searches and studies of its properties at the LHC
experiments due to its high resolution, small background and a clear
discrepancy between the measured signal strength as reported by
ATLAS \cite{atlas-2013,atlas-2013b} and CMS Collaboration
\cite{Roeck-lp2013}:
\beq
R_{\gamma\gamma}&=& 1.55\pm 0.23(stat)\pm 0.15(syst), ~~~~(ATLAS), \\
R_{\gamma\gamma}&=& 0.77\pm 0.27 ~~~~(CMS).
\eeq
Both measurements are still consistent with the SM prediction ($R_{\gamma\gamma}=1$) in
the 2$\sigma$ range at present due to still large errors. If the
excess (deficit) seen by ATLAS (CMS) is eventually confirmed by the
near future LHC measurements, the extra contributions from various
new physics (NP) models beyond the SM maybe help to understand such
excess or deficit. Of course, all extensions of the SM have to abide
by the existence of a Higgs boson with mass of about 125 GeV and
with SM-like properties.

The twin Higgs mechanism has been proposed as an alternative solution to the little hierarchy problem \cite{ly-1,ly-2}.
The idea of twin Higgs shares the same origin with that of little Higgs in
that the SM-like Higgs emerges as a pseudo-Goldstone boson \cite{little}. But rather than using collective symmetry breaking,
the twin Higgs mechanism takes an additional discrete symmetry to
stabilize the Higgs mass. The twin Higgs mechanism can be implemented in left-right Higgs (LRTH) model
with the discrete symmetry being identified with left-right symmetry \cite{ly-2}.
The phenomenology of the LRTH model has been extensively studied for example in Refs. \cite{Hock,dong,sf}.

The LHC diphoton signal has been studied in various new physics
models, such as some popular supersymmetry models \cite{susy}, the
two Higgs doublet model \cite{2hdm}, the Higgs triplet model
\cite{HTM}, the models with extra-dimensions \cite{ED}, the little
Higgs models \cite{LH}, and the other extensions of Higgs models
\cite{other1,other2}. In the LRTH model, the diphoton decay of the
SM-like Higgs boson was studied even before the LHC Higgs data
\cite{lrth1}. In this work, motivated by the latest LHC discrepancy
of $R_{\gamma\gamma}$, we will assume a SM-like Higgs boson with
125.5 GeV mass and study its implication in the LRTH model. Also we
will study some exclusive signal rates compared with the Higgs data
as well as the SM predictions. Besides, we will perform a global fit
to the latest LHC Higgs data to figure out if the LRTH model can
provide a better fit than the SM.

This paper is organized as follows. In the next section, we recapitulate the LRTH model and lay out the
couplings of the particles relevant to our calculation. In Sec. III, we investigate the LRTH model predictions
for the Higgs signal rates in light of the latest LHC experimental data.
Finally, we give our conclusion in Sec.IV.

\section{Relevant Higgs couplings in the LRTH model}\label{sec:intro}

The LRTH model is based on the global symmetry $U(4)\times U(4)$
with a locally gauged $SU(2)_{L}\times SU(2)_{R}\times U(1)_{B-L}$
subgroup. The twin symmetry is identified as the left-right symmetry
which interchanges L and R, implying that the gauge couplings of
$SU(2)_{L}$ and $SU(2)_{R}$ are identical $(g_{2L}=g_{2R})$. Two
Higgs fields, $H$ and $\hat{H}$, are introduced and each transforms
as $(4,1)$ and $(1,4)$ respectively under the global symmetry, which
can be written as
\beq H=\left( \begin{array}{c} H_{L}\\ H_{R} \\
\end{array}  \right)\,, \qquad
\hat{H}=\left( \begin{array}{c} \hat{H}_{L}\\ \hat{H}_{R} \\
\end{array}  \right)\,,
\eeq
where $H_{L,R}$ and $\hat{H}_{L,R}$ are two component objects
which are charged under the $SU(2)_{L}\times SU(2)_{R}\times
U(1)_{B-L}$ as
\beq H_{L}~and~ \hat{H}_{L}: (2, 1, 1),~~~~~~~~H_{R}~
and~ \hat{H}_{R}: (1, 2, 1).
\eeq
The global $U(4)_{1}(U(4)_{2})$ symmetry is spontaneously broken down to its subgroup
$U(3)_{1}(U(3)_{2})$ with non-zero vacuum expectation values(VEV) as
\beq
<H>=\left( \begin{array}{c} 0\\ 0\\ 0\\ f\\
\end{array} \right), \qquad <\hat{H}>=\left( \begin{array}{c} 0\\ 0\\ 0\\ \hat{f}\\
\end{array}\right).
\eeq
The Higgs VEVs also break $SU(2)_{R}\times U(1)_{B-L}$
down to the SM $U(1)_{Y}$. The details of the LRTH model as well as
the gauge sector, the fermion sector and Higgs sector have been
given in Ref.\cite{Hock}. Here we will focus on the new particles
and the couplings relevant to our work.

In the LRTH model, the heavy new gauge bosons ($W_{H}^{\pm},Z_{H}$),
heavy top quark partner ($T$) and other Higgs particles
($\phi^{0,\pm}$) are introduced to cancel the Higgs boson one-loop
quadratic divergence contributed by the gauge bosons, top quark and
Higgs boson of the SM. The masses of the particles that run in the
triangle loop diagrams are given in Ref.~\cite{Hock}. The relevant
Higgs couplings and the mixing angles for left-handed and
right-handed fermions are the following \cite{Hock}
\beq
{\mathcal{L}}&=& -\frac{m_{t}}{v}y_{t}\bar{t}th-\frac{m_{T}}{v}y_{T}\bar{T}Th+2\frac{m^2_{W}}{v}y_{W}W^{+}W^{-}h \nonumber\\
 &&
 +2\frac{m^2_{W_{H}}}{v}y_{W_{H}}W_{H}^{+}W_{H}^{-}h+2\frac{m^2_{Z}}{v}y_{Z}ZZh-2\frac{m_{\phi}^{2}}{v}y_{\phi}\phi^{+}\phi^{-}h,
 \\
s_{L}&=& \frac{1}{\sqrt{2}}\sqrt{1-(y^{2}f^{2}\cos2x+M^{2})/N_{t}},\\
s_{R}&=& \frac{1}{\sqrt{2}}\sqrt{1-(y^{2}f^{2}\cos2x-M^{2})/N_{t}},
\eeq where
$N_{t}=\sqrt{(M^{2}+y^{2}f^{2})^{2}-y^{4}f^{4}\sin^{2}2x}$ with
$x=v/\sqrt{2}f$ and $v=246$GeV is the electroweak scale, while $M$
is the mass parameter essential to the mixing between the SM-like
top quark and the heavy top quark. The explicit expressions of the
relevant couplings $ y_{t},  y_{T}, y_{W}, y_{W_{H}}$ and $y_{\phi}$
can be found easily in Ref.~\cite{Hock}.

 In the LRTH model, the relation between $G_{F}$ and $v$ is modified from its SM form, introducing an
 additional correction $y_{G_{F}}$ as $1/v^{2}=\sqrt{2}G_{F}y^{2}_{G_{F}}$ with $y^{2}_{G_{F}}=1-v^{2}/(6f^2)$.
 This correction must
 also be taken into account when comparing SM-like Higgs boson decay rates (i.e. $h \to XX$)
 in the LRTH model to the SM predictions with $G_{F}$ as input.

\section{Higgs decays in  the LRTH Model}

\subsection{The rates of $\sigma(gg \rightarrow h\rightarrow XX)$ at the LHC}

The Higgs production rates in the LRTH model normalized to the
SM values are generally defined as
\beq
R_{XX}=\frac{\sigma(pp \rightarrow h)Br(h\rightarrow XX)}{\sigma_{SM}(pp \rightarrow
h)Br_{SM}(h\rightarrow XX)},
\eeq
where $XX$ denotes $\gamma\gamma$, $Z\gamma$, $ZZ^{*}$, $WW^{*}$, or the SM fermion pairs.

At the LHC, the Higgs single production is dominated by the gluon-gluon fusion process. The hadronic cross section $\sigma(gg\rightarrow h)$ at leading order can be written as:
\beq
\sigma(gg\rightarrow h)&=& \frac{\pi^{2}\;
\tau_0}{8m_{h}^{3}}\Gamma(h\rightarrow gg)
\int^{1}_{\tau_{0}}\frac{dx}{x}f_{g}(x,\mu_{F}^{2})f_{g}(\frac{\tau_{0}}{x},\mu_{F}^{2}),
\eeq
where $\tau_{0}=m_{h}^{2}/s$ with $\sqrt{s}$ being the
center-of-mass energy of the LHC and $f_{g}(x,\mu_F^2)$ is the
parton distribution of gluon. Thus, one can see that the
$\sigma(gg\rightarrow h)$ has a strong correlation with the decay
width $\Gamma(h\rightarrow gg)$. Other main production processes of
the Higgs boson include vector-boson fusion (VBF), and associated
production with SM gauge bosons (VH) and top pair $t\bar{t}h$. For
$m_{h}=125.5$ GeV, the uncertainty on Higgs production has been
studied systematically by the LHC Higgs cross section working group
for the various channels and can be found easily in Ref.~\cite{handbook}.
The major decay modes of the Higgs boson are $h\rightarrow
f\bar{f}$($f=b,c,\tau$), $VV^{*}(V=W, Z)$, $gg$, $\gamma\gamma$ and
$Z\gamma$, where $W^{*}/Z^{*}$ denoting the off-shell charged or
neutral electroweak gauge bosons. The corresponding expressions are given in the Appendix.

The SM input parameters relevant in our study are taken from \cite{data}. The free LRTH model parameters involved are $f$, $M$,
and the masses of the charged Higgs bosons. The indirect constraints on $f$ come
from the $Z$-pole precision measurements, the low energy neutral current process and high energy precision
measurements off the $Z$-pole, requiring approximately $f> 500$ GeV. On the other hand, it cannot be too
large since the fine tuning is more severe for large $f$. The mixing parameter $M$ is constrained by the
$Z\rightarrow b\bar{b}$ branching ratio and oblique parameters.
Following Ref.~\cite{Hock}, we take the typical parameter space as:
\beq
500 GeV \leq f \leq 1500 GeV, \quad  0  \leq M \leq 150 GeV,
\eeq
while the mass $m_\phi$ of the charged Higgs boson $\phi^{\pm}$ is in the range of a few hundred GeV.

\begin{table}[thb]
\begin{center}
\caption{ The relative strength of the contributions to the decay
amplitude from various sources for $h\to \gamma\gamma$ and $h\to gg$
(numbers in the brackets )  in the SM and the LRTH model, assuming
$m_{\phi}=200$ GeV, $M=150$ GeV and $f=500,700,900,1100,1500$ GeV,
respectively. } \label{table2} \vspace{0.2cm}
\begin{tabular}{c|c|c|c|c|c|c} \hline\hline $m_{h}$=125.5 GeV&SM top&$W^\pm$&T-quark&$W_{H}$&$\phi^{\pm}$&total\\ \hline
SM&-1.84 (0.69)&8.34 (0)&0&0&0&6.50 (0.69) \\ \hline $f$=500
GeV&-1.68 (0.63)&7.84 (0)&0.18 (-0.07)&-0.031&-0.009&6.31 (0.56) \\
\hline $f$=700 GeV&-1.77 (0.66)&8.08 (0)&0.10
(-0.04)&-0.016&-0.004&6.40 (0.62) \\ \hline $f$=900 GeV&-1.79
(0.67)&8.19 (0)&0.06  (-0.024)&-0.01&-0.003&6.44 (0.65) \\ \hline
$f$=1100 GeV&-1.81 (0.68)&8.24 (0)&0.04 (-0.016)&-0.007&-0.002&6.46
(0.66)\\ \hline $f$=1500 GeV&-1.82 (0.68)&8.28 (0)&0.02
(-0.01)&-0.004&-0.001&6.48 (0.67)
\\  \hline\hline \end{tabular}\end {center} \end{table}

For the considered $h \to XX$ decays, one can write the decay amplitude $\cala(h\to XX)$ as the summation of
the pieces $\cala_i$ from different sources:
\beq
\cala(h\to XX) = \sum_{i=1}^{N} \cala_i(h \to XX). \label{eq:ai00}
\eeq
In Table I, we list all possible contributions to  the decay amplitude  $\cala(h\to  \gamma \gamma)$ and
$\cala(h\to  gg)$ coming from various sources, here we show the relative strength of different pieces only.

For the $h \to \gamma\gamma$ decay, for example, the SM contribution include two parts:  one comes
from the top quark loop with $\cala_{top}=-1.84$, another from the $W^\pm$ boson with
$\cala_{W}=8.34$. These two contributions have different sign and therefore interfere destructively.
In the LRTH model, however, the Feynman diagrams involving the $T-$quark, $W_H$ boson and $\phi^\pm$ boson
also provide the additional contributions to the decay $h \to \gamma\gamma$ respectively,
as illustrated explicitly in the column four to six of Table I.
From Table I we have the following observations:
\begin{enumerate}
\item
In the SM, the decay $h\to gg$ is dominated by the top quark loop,
while the contributions to $h\to \gamma\gamma$ arise from both the top quark
and $W$ boson loops simultaneously. The total decay amplitude of $h\to \gamma\gamma$ is clearly
dominated by the large positive contribution from the SM $W^\pm$ bosons loop.

\item
In the LRTH model, the additional new physics contributions are indeed much smaller in size than the SM part and
therefore play a minor role for the considered decay modes.

\item
Among the three NP sources, the contribution from the T-quark is the largest piece of the NP contributions,
but it is still too small to counteract with the positive SM part, this is because
the coupling $y_{T}$ is much smaller than $y_t$.  The NP contributions from $W_H$ and $\phi^\pm$
are even much smaller than the small T-quark piece and can be neglected safely.

\item
The NP contributions become smaller rapidly when $f$ becomes larger.
For $h \to \gamma\gamma$ decay, for example, the contribution from
the T-quark is changing from $0.18$ to $0.02$ when the parameter $f$
increases from 500 GeV to 1500 GeV.

\end{enumerate}

\begin{figure}[thb]
\begin{center}
\vspace{-0.5cm} \centerline{\epsfxsize=8cm\epsffile{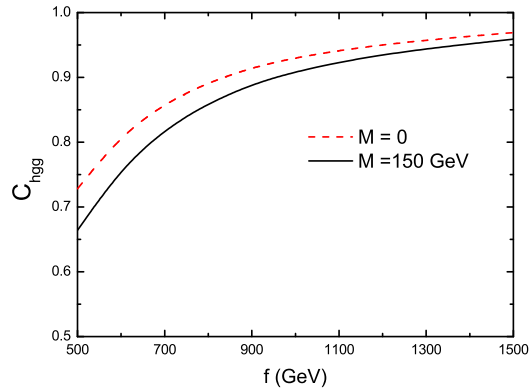}} \caption{
$f$-dependence of the ratio $C_{hgg}$ for two typical values of $M$
as indicated. } \label{fig:fig1}
\end{center}
\end{figure}

\begin{figure}[thb]
\begin{center}
\vspace{-0.5cm}
\centerline{\epsfxsize=8cm\epsffile{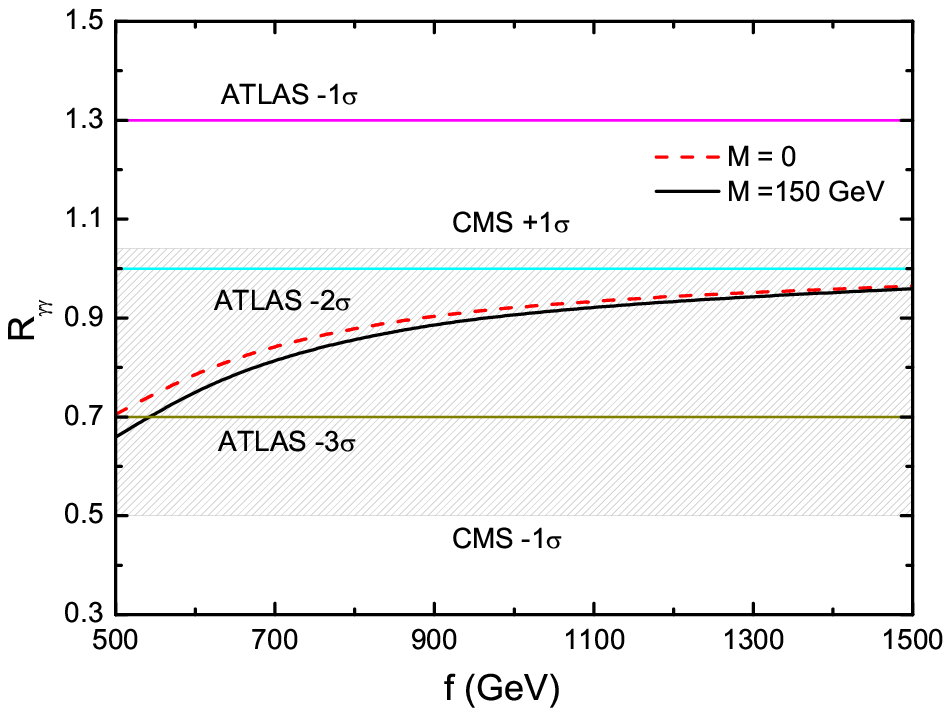}
\hspace{-0.5cm}\epsfxsize=8cm\epsffile{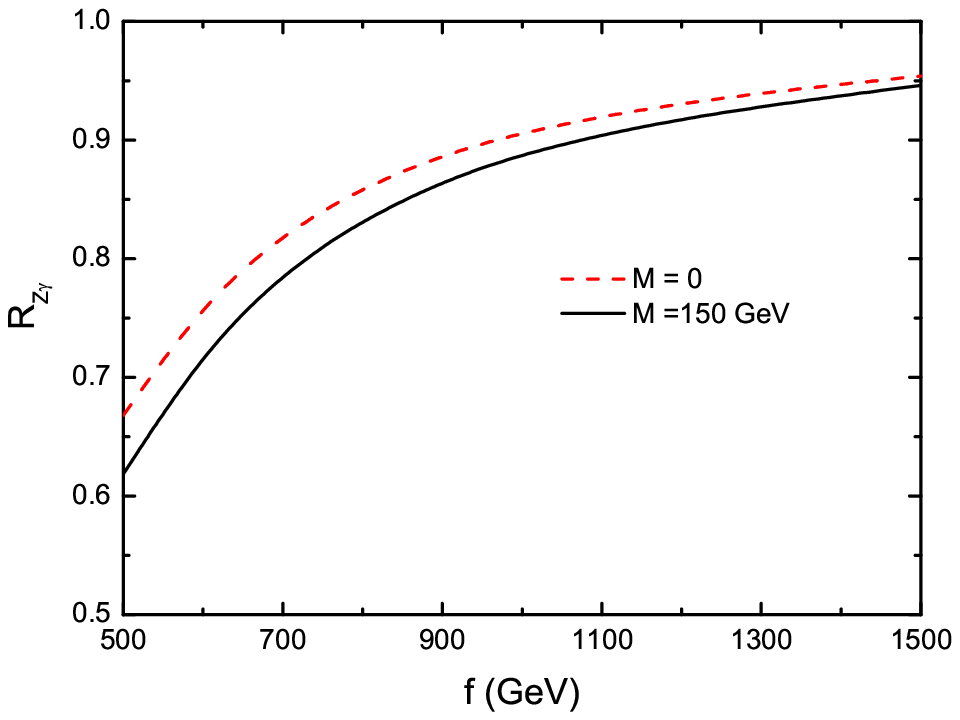}}
\caption{$f$-dependence of $R_{\gamma\gamma}$(left) and $R_{Z\gamma}$(right) for two typical values of $M$ as indicated.
The shaded area shows the CMS result: $R_{\gamma\gamma}=0.77\pm 0.27$.}
\label{fig:fig2}
\end{center}
\end{figure}

\begin{figure}[thb]
\begin{center}
\vspace{-0.5cm}
\centerline{\epsfxsize=8cm\epsffile{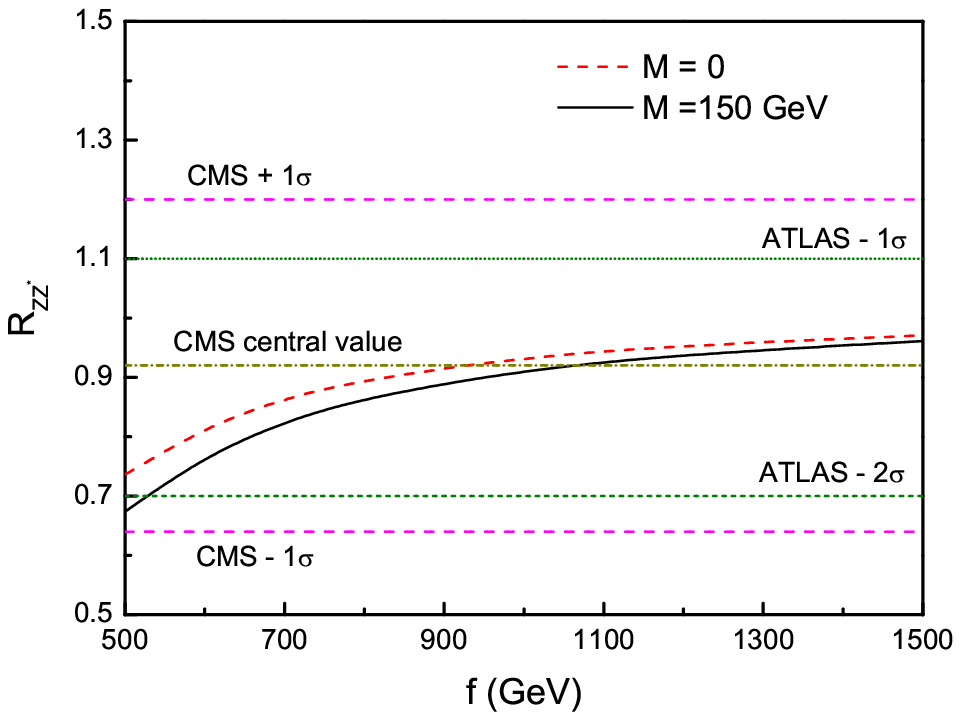}
\hspace{-0.5cm}
\epsfxsize=8cm\epsffile{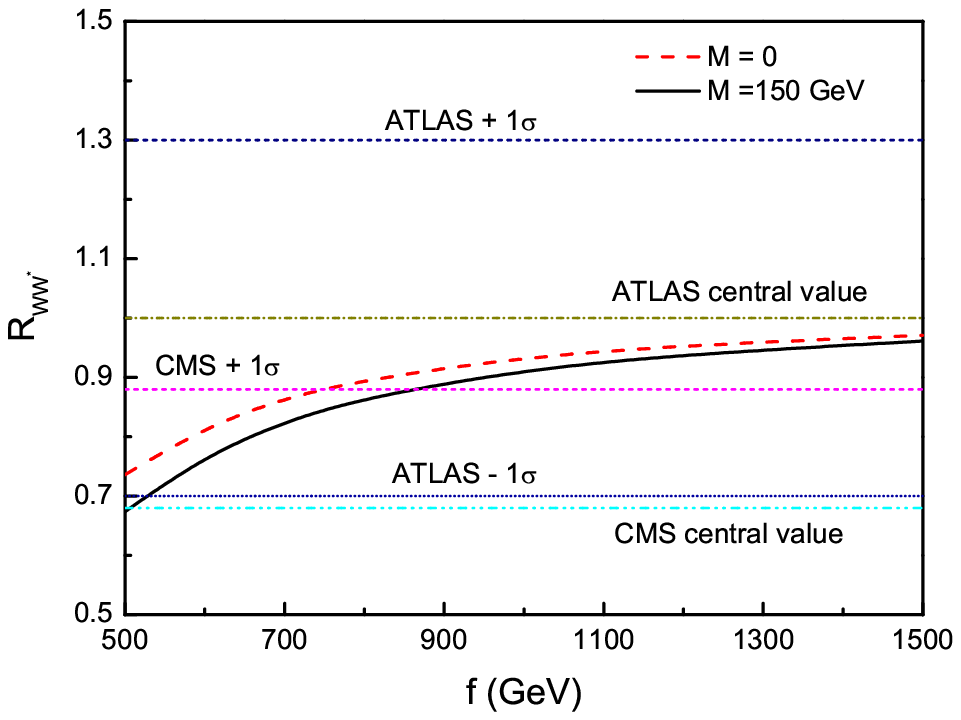}}
\caption{ $f$-dependence of $R_{ZZ^{*}}$(left) and $R_{WW^{*}}$(right) for two typical values of $M$ as indicated. }
\label{fig:fig3}
\end{center}
\end{figure}

In Fig.1 we show the $f$-dependence of the ratios
$C_{hgg}=\Gamma_{LRTH}(h\to gg)/\Gamma_{SM}(h\to gg)$ for two
typical values of $M$: $M=0,150$ GeV. Here $\Gamma_{SM}(h\to gg)$
denotes the decay width of $h\to gg$ in the SM. One can see that the
NP correction becomes smaller rapidly along with the increase of the
parameter $f$, but becomes larger when $M$ is increasing. This is
because the parameter $M$ is introduced to generate the mass mixing
term $M q_{L}q_{R}$, and the LRTH model can give corrections via the
coupling of $ht\bar{t}$ and the heavy T-quark loop. For the special
case of $M=0$, there is no mixing between the SM top quark and the
heavy $T$ quark. By assuming $f=500$ GeV and varying $M$ in the
range of $0\leq M \leq 150$ GeV, the NP correction can be changed
from $17\%$ to $34\%$ to the SM value.

We know that the large experimental and theoretical uncertainties may  prevent
the detection of the deviation of the LRTH model prediction of $C_{hgg}$ from the SM one
for large value of scale $f$.
The QCD corrections to the total cross section of $h \to gg$
have been computed at next-to-next-to-leading order (NNLO) in Ref~.\cite{HK2002}.
The remaining renormalization/factorization scale
dependence of the cross section gives a lower bound on the size of the theoretical
uncertainty due to uncalculated higher-order QCD radiative corrections of about $15\%$
\cite{CS1101}, which can be further reduced with the inclusion of recently known
NNNLO results as described in Ref.~\cite{Ball13}.

In Fig.2 we plot the ratio $R_{\gamma\gamma}$ and $R_{Z\gamma}$ versus $f$ for two typical values of $M$ in the LRTH model.
It can be seen from Fig.2 that the ratio $R_{\gamma\gamma}$ and $R_{Z\gamma}$  in the LRTH model
are always smaller than unit, and will approach one for a large $f$.
On the other hand, for a small value of parameter $f$, the deviation from the SM prediction
is sensitive to the mixing parameter $M$.

For the diphoton signal, the measured value of  $R_{\gamma\gamma}=0.77\pm 0.27$ as reported by CMS Collaboration
can be understood in the LRTH model.
Of course, the LRTH prediction for  $R_{\gamma\gamma}$ is always outside 2$\sigma$ range of the ATLAS result.
The key point here is the large difference between the central values of the measured $R_{\gamma\gamma}$ as reported
by ALTAS and CMS Collaborations. Further improvement of the $R_{\gamma\gamma}$ measurements for both ATLAS and CMS
Collaboration is greatly welcome and will play the key role in constraining the new physics models beyond the SM.

For the $h\to Z\gamma$ channel there is not enough data to draw any conclusion about LRTH.
For the ratios  $R_{ZZ^*}$ and $R_{WW^*}$, the ATLAS and CMS measurements are consistent
with each other within one standard deviation.
In Fig.3 we plot the $f-$dependence of the ratio $R_{ZZ^*}$ and $R_{WW^*}$ for two typical values of $M$.
It can be seen from Fig.3 that the ratio  $R_{ZZ^*}$ and $R_{WW^*}$ in the LRTH model are
always smaller than unit and sensitive to the value of parameter $f$ and $M$.

\begin{table}[thb]
\begin{center}
\caption{ The theoretical predictions for the Higgs production rates
$R_{XX}$ in the LRTH model, assuming $m_{\phi}$=200 GeV, $M$=150 GeV
and $f=500,800,1200$ and $1500$ GeV. The corresponding measured
values reported by ATLAS and CMS
\cite{Roeck-lp2013,atlas-2013,atlas-2013b} are listed as comparison.}
 \label{table2a}
\vspace{0.2cm}
\begin{tabular}{c|c|c|c|c|c} \hline\hline $f$ (GeV)&$R_{\gamma\gamma}$&$R_{ZZ^{*}}$&$R_{WW^{*}}$&$R_{\tau^{+}\tau^{-}}$&$R_{Z \gamma}$\\
\hline 500&0.659&0.674&0.674&0.674&0.619 \\ \hline
800&0.858&0.866&0.866&0.866&0.833 \\ \hline
1200&0.936&0.939&0.939&0.939&0.92 \\ \hline
1500&0.959&0.961&0.961&0.961&0.946 \\ \hline
ATLAS&$1.55\pm0.23\pm0.15$&$1.43\pm0.33\pm0.17$&$0.99\pm0.21\pm0.21$&$0.7\pm 0.7$&$<13.5$
\\
\hline
CMS&$0.77\pm0.27$&$0.92\pm0.28$&$0.68\pm0.2$&$1.1\pm0.41$&$<9.3$ \\
 \hline \hline
\end{tabular} \end{center}\end{table}

In Table II, we list the LRTH predictions for the Higgs boson
production rates $R_{\gamma\gamma}, R_{WW^*}, R_{ZZ^*},R_{\tau\tau}$
and $R_{Z\gamma}$, assuming $M=150$GeV, $m_{\phi}=200$ GeV and $500
\leq f \leq 1500$ GeV. From the numerical results as listed in Table
II, one can see the five signal rates are always suppressed when
the new physics contributions are taken into account, which is
similar with the situation in the little Higgs models
\cite{like-lh}. This is mainly due to the following common reasons
in these kind of new physics models:
\begin{enumerate}
\item
The couplings of top quark partner $T$ and new heavy gauge bosons
$W_{H}$ with the Higgs boson have the opposite sign with respect to the Higgs couplings with SM top
quark and gauge bosons, respectively.

\item
The new physics part of the Higgs couplings to the SM top quark and gauge bosons are
suppressed by the ratio $v^{2}/f^{2}$, and will become zero in the limit
$f \to \infty$.

\end{enumerate}

It is well known that the production and decays of the Higgs boson are largely
affected by high order corrections. In order to reduce the errors
of theoretical predictions, we defined $R_{XX}$ as the ratios of the theoretical
predictions in the SM and in the LRTH model. In this way, the theoretical
errors will be largely canceled.

In many cases, the higher order corrections to the relevant cross sections or the
branching ratios could be factorized out approximately as simple factors (NLO, or NNLO, etc)
of the leading order results as discussed in Ref.~\cite{jhep13-029}.
For instance, one can see that the NLO QCD corrections to both $hgg$ and
$h\gamma\gamma$ vertex can give a simple multiplicative factor.
We assume that the QCD corrections in the LRTH model are similar as
those in the SM top loop for simplicity, thus the QCD corrections cancel to
a large extent in these ratios, provided that a single production mechanism
dominates. This certainly applies to $\mu_{\gamma\gamma}$, $\mu_{VV}$, and
$\mu_{\tau^{+}\tau^{-}}$ which are governed by the dominant production channel
through gluon fusion \cite{jhep2}.

\subsection{Global fit of the LRTH model to current LHC Higgs data}

By using the latest LHC Higgs data of 17 channels from both ATLAS
and CMS as given in Refs.~\cite{atlas-2013-1,cms-2013}, we now
perform a global fit to the LRTH model with the method proposed in
\cite{jhep1,jhep2}. When fitting the various observables, we
consider the correlation coefficients given in
Ref.\cite{jhep-1307-065} due to the independent data for different
exclusive search channels by two collaborations.

The global $\chi^{2}$ function is defined as usual:
\begin{equation}
\chi^{2}=\sum_{i,j}(\mu_{i}-\hat{\mu}_{i})(\sigma^{2})^{-1}_{ij}(\mu_{j}-\hat{\mu}_{j}),
\end{equation}
where index $i,j$ runs over all the different production/decay channels considered in this paper,
$(\mu_i,\mu_j)$ and $(\hat{\mu}_i, \hat{\mu}_j)$ are the corresponding theoretical
signal strength in the LRTH model and the measured Higgs signal strengths  as reported by both ATLAS and
CMS collaborations, respectively.
$\sigma^{2}_{ij}=\sigma_{i}\rho_{ij}\sigma_{j}$, $\sigma$ is
the experimental error extracted from the data at 1$\sigma$ and
$\rho_{ij}$ is the correlation matrix. Taking two correlated observables for instance, the correlation coefficient $\rho$ is applicable to the
following formula
\beq
\chi_{1,2}^{2}=\frac{1}{(1-\rho^{2})}\cdot
\left [\frac{[\mu_{1}-\hat{\mu}_{1}]^{2}}{\sigma_{1}^{2}}
+\frac{[\mu_{2}-\hat{\mu}_{2}]^{2}}{\sigma_{2}^{2}}-2\rho\frac{[\mu_{1}-\hat{\mu}_{1}]
\cdot[\mu_{2}-\hat{\mu}_{2}]}{\sigma_{1}\sigma_{2}} \right ].
\eeq
Note that the errors on the reported Higgs signal strengths
$\hat{\mu}_{i}$ are symmetrized by the relation
\beq
\delta\hat{\mu}_{i}=\sqrt{[(\delta\hat{\mu}_{+})^{2}+(\delta\hat{\mu}_{-})^{2}]/2},
\eeq
where $\delta\hat{\mu}_{\pm}$ are the one-sided errors given by the
experimental collaborations. For plotting distributions of a
function of one variable, the $68\%$ ($1\sigma$) and $95\%$
($2\sigma$) confidence level (CL) intervals are obtained by
$\chi^{2}=\chi_{min}^{2}+1$ and $+4$, respectively.
For a more detailed description of the fit procedure, one can see
Refs.~\cite{jhep1,jhep2,jhep-1307-065}.

\begin{figure}[t]
\begin{center}
\vspace{-0.5cm} \centerline{\epsfxsize=9cm\epsffile{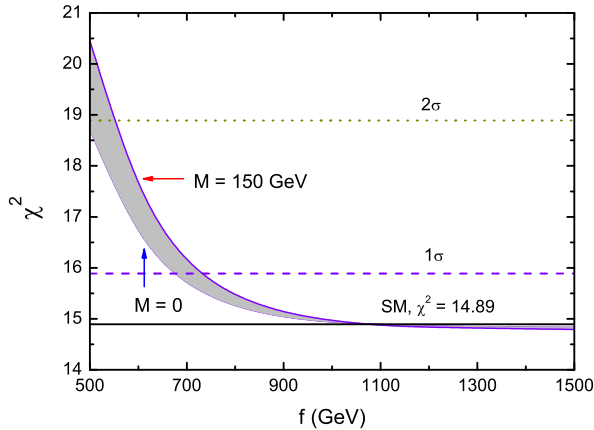}} \caption{
The global fit values of $\chi^{2}$ versus $f$ for $M=0$ and $150$
GeV.} \label{fig:fig4}
\end{center}
\end{figure}

In Fig. 4 we project the samples on the global fit values of
$\chi^{2}$ versus parameter $f$ for $ M=0$ and $150$ GeV. One can
see that the value of $\chi^{2}$ is larger than that for SM for most
of parameter space of $f$ and approaches the SM value for a
sufficiently large $f$. For a large values of scale $f$ (about 1100
GeV), it is slightly smaller than the SM value ($\chi^{2}=14.88$ for
$M=150$ GeV while $\chi_{SM}^{2}=14.89$).
So we can see that the good points favored by the current LHC Higgs data is
at the region of $f\geq 1100$ GeV. For $M=150$ GeV and $f < 550$ GeV, the value
of $\chi^2$ is larger than $18.9$, which implies that $f < 550$ GeV is excluded
at $95\%$ confidence level from the experimental viewpoint

\begin{figure}[t]
\begin{center}
\vspace{-0.5cm} \centerline{\epsfxsize=14cm\epsffile{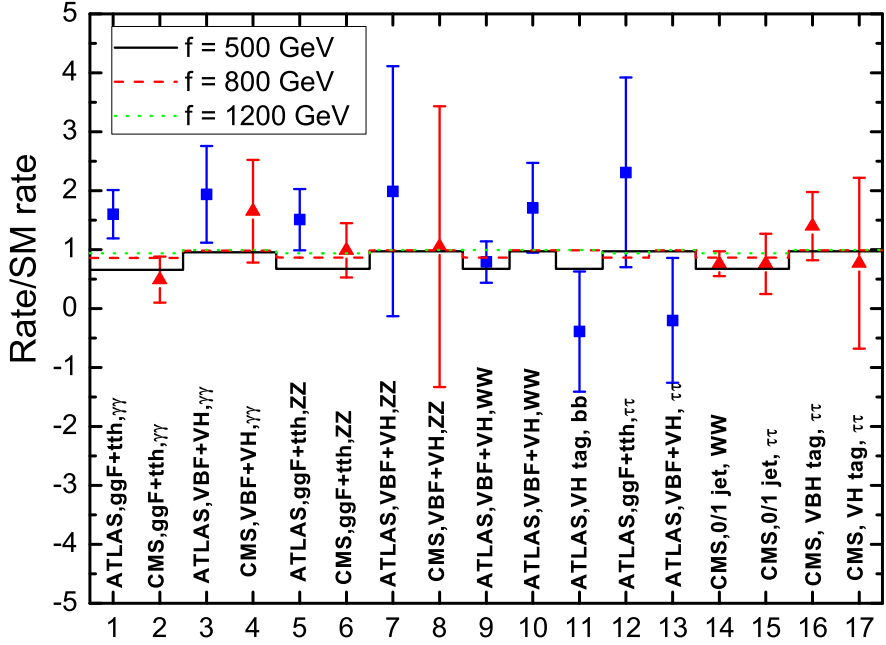}}
\vspace{-1cm} \caption{ The LRTH predictions for the various Higgs
signal rates $R_{XX}$ at the LHC, assuming $M=150$ GeV, $f=500, 800$
and $1200$ GeV respectively. The error-bars show the ATLAS and CMS
measurements of 17 channels as given in
Refs.~\cite{atlas-2013-1,cms-2013}. } \label{fig:fig5}
\end{center}
\end{figure}

In Fig.5 we present the LRTH predictions of different Higgs signal
rates $R_{XX}$, and a comparison with the corresponding experimental
measurements at the LHC, assuming  $M=150$ GeV and the scalar
parameter $f=500, 800$ and $1200$ GeV  respectively. In our fit, we
select 17 sets of data from Refs.\cite{atlas-2013-1,cms-2013}.
From Fig.5 one can see that all the signal rates are suppressed due to
the inclusion of new physics corrections in the LRTH model, when
compared with the SM values. In the LRTH model, we find
$\chi^2=20.29,15.39, 14.82$ for $f=500, 800$ and $1200$ GeV. The
LRTH prediction for $R_{\gamma\gamma}$ agree well with the CMS
measurement: $R_{\gamma\gamma}^{\rm CMS}=0.77\pm 0.27$.

For given values of the LRTH parameter $M$ and $f$, the masses $M_T$, $M_{W_H}$ and the relevant couplings
$y_t,y_T$ and $y_{W}$ will be determined consequently.
In Table III we present the numerical results of the LRTH predictions for some ratios and various
Higgs signal rates, as illustrated explicitly in Fig.~5.

\begin{table}[thb]
\begin{center}
\caption{ The numerical results of the LRTH predictions for some ratios and various
Higgs signal rates, assuming $M=0, 150$ and $f=500$, $800$ GeV, respectively. }
 \label{table3}
\vspace{0.2cm}
\begin{tabular}{|c|cc|cc|} \hline\hline $M$ (GeV) & \multicolumn{2}{c|}{0}&\multicolumn{2}{c|}{150}\\ \hline
$f$ (GeV) &500 &800&500 &800 \\ \hline \hline
 $m_{T}$ (GeV)&464.9&774.4&488.5&788.8 \\ \hline
 $m_{W_{H}}$ (GeV)&1175.6&1883.7&1175.6&1883.9 \\ \hline
 $y_{t}^{2}$ &1.0&1.0&0.871&0.959 \\ \hline
 $y_{T}^{2}$ &0.017&0.002&0.011&0.002 \\ \hline
 $y_{W}^{2}$ &0.921&0.969&0.921&0.969 \\ \hline \hline
  $C_{hgg}$&0.728&0.892&0.664&0.861 \\ \hline
 $C_{h\gamma\gamma}$&0.919&0.966&0.939&0.976 \\ \hline
 $C_{hZ\gamma}$&0.871&0.944&0.881&0.947 \\ \hline
 $C_{hVV^{\ast}}$&0.921&0.969&0.921&0.969 \\ \hline
ggF+ttH, $\gamma\gamma$&0.705 &0.882&0.663&0.858 \\
\hline VBF+VH, $\gamma\gamma$&0.931&0.971&0.953&0.982 \\
 \hline
 ggF+ttH, $ZZ$&0.736 &0.896&0.674&0.866 \\
\hline VBF+VH, $ZZ$&0.971 &0.989&0.974&0.989 \\
\hline
 ggF+ttH, $WW$&0.736 &0.896&0.674&0.866 \\
\hline VBF+VH, $WW$&0.971 &0.989&0.974&0.989 \\
\hline
 VH tag, $b\bar{b}$&0.971 &0.989&0.974&0.989 \\
 \hline
 ggF+ttH, $\tau\tau$&0.736 &0.896&0.674&0.866 \\
\hline VBF+VH, $\tau\tau$&0.971 &0.989&0.974&0.989 \\
\hline
 0/1 jet, $WW$&0.973 &0.897&0.674&0.866 \\
\hline 0/1 jet, $\tau\tau$&0.941 &0.898&0.681&0.868 \\
\hline VBF tag, $\tau\tau$&0.998 &0.999&1.000&1.001 \\
\hline VH tag, $\tau\tau$&0.971 &0.989&0.974&0.989 \\ \hline \hline
 $\chi^{2}$&18.55&15.19&20.3&15.39\\ \hline
\hline
\end{tabular} \end{center}\end{table}

In the near future, the improved measurement of the diphoton signal at the LHC
will play a decisive role for these models.
For example, if the future well-measured diphoton rate is still clearly larger than unit,
the LRTH model and other little Higgs models will be strongly disfavored or ruled out.
Otherwise, if the deficit signal rate permits, these models will be favored.
However, it is difficult for the LHC to clearly discriminate these
new physics models due to the different free parameters for each model.
The high energy and high luminosity linear electron positron collider experiments, such as CLIC or
the ILC, will provide a rather clean environment for new physics discovery \cite{ilc}.

\section{ Conclusions}

In this work, we studied the Higgs production and decay in the LRTH model in the light
of the latest LHC Higgs data from ATLAS and CMS Collaboration.
From the numerical results we obtain the following observations:
\begin{enumerate}
\item
The signal rates normalized to the SM prediction for the five Higgs search channels
are always suppressed when new physics contributions are taken into account and
approach the SM predictions for a large scale parameter $f$.

\item
The LRTH prediction for $R_{\gamma\gamma}$  agree well with the CMS measurement at $1\sigma$ level,
but differ with the ATLAS result. The LRTH model could be further tested by the
improved measurement of $R_{\gamma\gamma}$ at LHC.

\end{enumerate}

\begin{acknowledgments}

This work is supported by the National Natural Science
Foundation of China under the Grant No. 11235005 and the Joint Funds
of the National Natural Science Foundation of China (U1304112).

\end{acknowledgments}


\begin{appendix}

\section{The Higgs decays in the LRTH model }\label{sec:higgs}

In the LRTH model, the decays $h\rightarrow gg, \gamma\gamma, Z\gamma$ all receive contributions
from the modified couplings $hXX$ and the new heavy particles. The LO decay widths of $h\rightarrow
 gg, \gamma\gamma, Z\gamma$ are given by
\beq
 \Gamma(h\to gg)&=&\frac{\sqrt{2}G_{F}\alpha_{s}^{2}m_{h}^{3}}{32\pi^{3}}
 \Bigl |-\frac{1}{2}F_{1/2}(\tau_{t})y_{t}y_{G_{F}}-
 \frac{1}{2}F_{1/2}(\tau_{T})y_{T} \Bigr |^{2},\\
 \Gamma(h\to  \gamma\gamma)&=&\frac{\sqrt{2}G_{F}\alpha_{e}^{2}m_{h}^{3}}{256\pi^{3}}
 \Bigl | \frac{4}{3}F_{1/2}(\tau_{t})y_{t}y_{G_{F}}+\frac{4}{3}F_{1/2}(\tau_{T})y_{T}\non
 & &
 +F_{1}(\tau_{W})y_{W}+F_{1}(\tau_{W_{H}})y_{W_{H}}+F_{0}(\tau_{\phi})y_{\phi} \Bigr |^{2},\\
\Gamma(h\to  Z\gamma)&=&\frac{\alpha_{e}^{2}m_{h}^{3}}{128\pi^{3}s^{2}_{W}c^{2}_{W}v^{2}}
\left (1-m_{Z}^{2}/m_{h}^{2} \right )^{3}\non
&& \cdot \left | 2y_{f}(1-\frac{8}{3}s_{W}^{2})A_{1/2}(\tau_{f},\lambda_{f})
+y_{W}c^{2}_{W}A_{1}(\tau_{W},\lambda_{W}) \right |^{2},
\eeq
with
\beq
 F_{1}&=&2+3\tau+3\tau(2-\tau)f(\tau),\non
F_{1/2}&=&-2\tau[1+(1-\tau)f(\tau)],\non
F_{0}&=&\tau[1-\tau f(\tau)],\non
A_{1}&=&4(3-\tan^{2}\theta_{W})I_{2}(\tau,\lambda)
+(1+2\tau^{-1})\tan^{2}\theta_{W}-(5+2\tau^{-1})I_{1}(\tau,\lambda),\non
A_{1/2}&=&I_{1}(\tau,\lambda)-I_{2}(\tau,\lambda),
\eeq
where
\beq
I_{1}(\tau,\lambda)&=&\frac{\tau\lambda}{2(\tau-\lambda)}+\frac{\tau^{2}\lambda^{2}}{2(\tau-\lambda)^{2}}[f(\tau)-f(\lambda)]+\frac{\tau^{2}\lambda}{(\tau-\lambda)^{2}}[g(\tau)-g(\lambda)],\\
I_{2}(\tau,\lambda)&=&-\frac{\tau\lambda}{2(\tau-\lambda)}[f(\tau)-f(\lambda)],
\eeq
with
\beq
f(\tau)&=&\left [\sin^{-1}(1/\sqrt{\tau}) \right ]^{2},\non
g(\tau)&=& \sqrt{\tau-1}\sin^{-1}(1/\sqrt{\tau}),
\eeq
for $\tau_{i}=4m_{i}^{2}/m_h^2 \geq  1$.

The partial decay widths into single off-shell gauge bosons $h\rightarrow
VV^{*}$ are given in Ref.~\cite{zz}
\beq
\Gamma(h\to WW^{*})&=&\frac{3G_{F}^{2}m_{W}^{4}m_{h}}{16\pi^{3}}
F\left (\frac{m_{W}^{2}}{m_{h}^{2}} \right ), \\
\Gamma(h\to ZZ^{*})&=&\left ( \frac{7}{4}-\frac{10}{3}s_{W}^{2}+\frac{40}{9}s_{W}^{4}
\right)\frac{G_{F}^{2}m_{Z}^{4}m_{h}}{16\pi^{3}}F\left (\frac{m_{Z}^{2}}{m_{h}^{2}} \right ),
 \eeq
with the form factor $F(x)$ is formulated as
\beq
 F(x)&=&\frac{x-1}{2x}\left ( 2-13x+47x^{2} \right)
 -\frac{3}{2}\left ( 1-6x+4x^{2} \right )\ln x \non
 & &
 +\frac{3(1-8x+20x^{2})}{\sqrt{4x-1}}\arccos\left (\frac{3x-1}{2x^{3/2}} \right).
\eeq

\end{appendix}


\end{document}